\documentclass[sigconf]{acmart}
\AtBeginDocument{%
  }

\copyrightyear{2025}
\acmYear{2025}
\acmDOI{}
\acmISBN{}

\acmConference[AlpCHI '26, WS2: Human Cognition, AI, and the Future of HCI]{Navigating the Disruptive and Wild Landscape of Large Language Models and Agentic AI}{Mar. 1--5, 2026}{Ascona, Switzerland}

\usepackage{tikz}
\usetikzlibrary{shapes, arrows.meta, positioning, calc, shadows}

\usepackage{tikz}

\newcommand\submittedtext{%
  \footnotesize \center \textbf{Paper selected for the workshop \textit{Human Cognition, AI, and the Future of HCI: Navigating the Disruptive and Wild Landscape of Large Language Models and Agentic AI} as part of the \textit{Human-Computer Interaction (HCI) conference of the Alpine region} (AlpCHI 2026) hosted at the Congressi Stefano Franscini, March 1st to March 5th, 2026 on Monte Verità in Ascona, Switzerland.}}

\newcommand\submittednotice{%
\begin{tikzpicture}[remember picture,overlay]
\node[anchor=south,yshift=20pt] at (current page.south) {
  \parbox{\dimexpr0.85\textwidth-\fboxsep-\fboxrule\relax}{\submittedtext}
};
\end{tikzpicture}%
}

\begin{document}

\title[From Particles to Agents]{From Particles to Agents: Hallucination as a Metric \\ for Cognitive Friction in Spatial Simulation}


\author{Javier {Argota Sánchez-Vaquerizo}}
\authornote{Both authors contributed equally to this research.}
\email{javier.argota@gess.ethz.ch}
\affiliation{%
  \institution{Computational Social Science. ETH Zürich}
  \city{Zurich}
  \country{Switzerland}
}

\authornote{Corresponding author.}

\author{Luis {Borunda Monsivais}}
\authornotemark[1]
\email{lborunda@vt.edu}
\orcid{}
\affiliation{%
  \institution{Virginia Tech School of Architecture}
  \city{Blacksburg}
  \state{Virginia}
  \country{USA}
}


\begin{abstract}
  Traditional architectural simulations (e.g. Computational Fluid Dynamics, evacuation, structural analysis) model elements as deterministic physics-based "particles" rather than cognitive "agents". To bridge this, we introduce \textbf{Agentic Environmental Simulations}, where Large Multimodal generative models actively predict the next state of spatial environments based on semantic expectation. Drawing on examples from accessibility-oriented AR pipelines and multimodal digital twins, we propose a shift from chronological time-steps to Episodic Spatial Reasoning, where simulations advance through meaningful, surprisal-triggered events. Within this framework we posit AI hallucinations as diagnostic tools. By formalizing the \textbf{Cognitive Friction} ($C_f$) it is possible to reveal "Phantom Affordances", i.e. semiotic ambiguities in built space. Finally, we challenge current HCI paradigms by treating environments as dynamic cognitive partners and propose a human-centered framework of cognitive orchestration for designing AI-driven simulations that preserve autonomy, affective clarity, and cognitive integrity.

\end{abstract}




\maketitle

\submittednotice
\section{Architectural simulations are becoming agentic}



Current design, architectural and urban simulation paradigms, e.g. Computational Fluid Dynamics (CFD) for airflow, or physics-based for crowd evacuation, share a fundamental limitation: they model people as "particles" governed by deterministic physics rather than semantically-aware "agents" governed by malleable cognition.
However, we are experiencing a transition towards an Agentic Environmental Simulation, where the fundamental computational unit is not the vector but a "reasoning loop".
Recent work in large-scale social simulacra, such as the \textit{SocioVerse} framework \citep{Zhang2025SocioVerse:Users}, demonstrates that Large Language Model (LLM) agents can exhibit emergent social behaviors, memory retention, and adaptive planning that static rule-based systems cannot replicate\citep{Xi2025TheSurvey}.
By embedding these generative AI agents into spatial pipelines, we can move from simulating how a body collides with or avoids a wall to simulating how a mind collides and interfaces with a fuzzy spatial signal.
This allows for the prediction of complex, non-deterministic phenomena, such as dynamic crowd turbulence caused by ambiguous signage, that strictly physical simulations fail to capture.

\section{From Chronological Physics to Episodic Cognition}

Traditional architectural simulations operate on a chronological lockstep $(t+1)$, prioritizing high-frequency physical fidelity over an experiential reality. 
However, human spatial memory encodes the built environment through low-frequency semantic fidelity, structuring experience around discrete, meaningful events \citep{Zacks2007EventSegmentation}. We propose a shift to \textbf{Episodic Spatial Reasoning}, where simulation advances through "narrative beats" and cognitive boundaries (i.e. doorways) that trigger cognitive segmentation \citep{Radvansky2006WalkingSpace}.

By treating the environment as a structured semantic prompt chain, our pipeline mirrors dual-process cognition \citep{Kahneman2011Thinking} to optimize computational resource allocation: \begin{itemize} 
\item \textbf{Heuristic Autopilot (System 1):} A low-compute background layer handling routine locomotion via physics-based heuristics. 
\item \textbf{Episodic Reasoning (System 2):} High-compute Multimodal LLM modules activated only during critical episodes, such as spatial junctions, semiotic ambiguities, or affective transitions, i.e. when System 1 exceeds a surprisal threshold \citep{Zacks2007EventSegmentation}. 
\end{itemize}

This decoupling from exhaustive geometric surveying enables \textbf{Cognitive Simulations} that quantify qualitative experience previously restricted to a post-occupancy evaluation. 
These "empathy engines" generate heatmaps of affordance \citep{Qian2024AffordanceLLM:Models}, perception, and emotional loads (stress, risk, comfort). 
Given that LLMs can proxy human affective transitions when provided with rich contextual narratives \citep{Wang2023EmotionalModels, Hu2025AAgents}, this pipeline can simulate how specific neurodivergent individuals or socio-demographic profiles perceive space differently from normative standards.

The output is a multimodal narrative log of "Moments of Disorientation" and "Semiotic-Cognitive Misalignment". 
Crucially, this framework distinguishes between \textit{productive friction} orchestrated for reflection or awe, and \textit{cognitive hazardous friction} resulting from unintentional disorientation. 
It predicts, identifying and exploring \citep{ArgotaSanchez-Vaquerizo2025}, where an environment fails to support a person's cognitive intent, and therefore revealing hidden 'cognitive hazards' in the spatial design, i.e. the friction of a space.

This enables a layered analysis of spatial experience: (1) \textbf{Geometry} (the container), (2) \textbf{Navigability} (the path), (3) \textbf{Semantics} (the meaning), and (4) \textbf{Affect} (the feeling). Ultimately, this shifts the design objective from utilitarian optimization (flow, thermal comfort) to \textbf{cognitive orchestration}, ensuring the environment supports the occupant's mental integrity as robustly as its structure supports their physical weight.

\section{Not a bug, but a feature: Hallucination as Design Heuristic}

We propose an alternative interpretation of generative error, so-called hallucinations, as design heuristic \citep{2022MachineIntelligence}.
Fundamentally, LLMs lack an internal "ground truth" World Model \citep{LeCun2022A0.9.2}.
Instead, they rely solely on probabilistic associations \citep{Xu2024HallucinationModels}.
Conventional approaches toward generative AI validation penalize models that deviate from ground truth. 
While this precision is essential to accurately simulate physical environments \citep{Borunda2025ANNSIM}, we argue that in the context of human-environment interaction, specific types of AI hallucinations serve as valid and operationalizable proxies for structural semiotic ambiguity, identifying where environmental signal fails to clearly communicate affordances.
Hence, these "hallucination hotspots" do not indicate model failure, but rather reveal ambiguous architectural semiotics, i.e. areas where the built environment is likely to confuse, stress, or mislead some people.

For instance, when a Vision-Language Model (VLM) incorrectly identifies a glass partition as an open passage, it can be related to the same high-frequency visual noise and semantic ambiguity that triggers "pareidolia" or false spatial affordance observed in human perception.
Recent work on programmable cognitive biases \citep{Liu2025ProgrammableAgents} shows how it is possible to replicate biases and "bounded rationality".
This extends AI-driven AR accessibility pipelines \citep{Borunda2025ANNSIM} to expose not just physical bottlenecks (e.g. narrow corridors) but semiotic spatial bottlenecks (e.g. areas where the design is so ambiguous that it makes the agent "hallucinate").
By mapping these hallucination hotspots, it is possible to generate a "Cognitive Friction Heatmap", identifying zones where the built environment results in ambiguity, where legibility \citep{Lynch1960} of intent may not be clear. In this context, Hallucination becomes a feature of subjective interpretation.

\subsection{Quantifying Cognitive Friction: The Hallucination-Reality Gap}
To operationalize hallucination as a heuristic, we formalize \textbf{Cognitive Friction} ($C_f$) as a metric of \textbf{Semiotic Divergence}. Whereas conventional simulation defines error as the agent’s deviation from geometric constraints ($A\neq R$), we invert this definition to characterize error as the environment's failure to transmit affordance ($R\centernot{\Rightarrow}A$). Thus, friction serves as a measure of \textit{semiotic ambiguity} occurring precisely where physical reality fails to clearly inform the agent's perceptual system.

For any spatial state $S$, the agent's generative model (System 2) produces an aggregated probabilistic expectation of the proximal state ($E_{gen}$), or "hallucination", across $N$ iterations. The physical environment provides the ground truth ($R_{phys}$). Cognitive Friction is the semantic distance between this expectation and reality:

\begin{equation} C_f = 1 - \text{sim}(E_{gen}, R_{phys}) \end{equation}

Where $sim$ is the cosine similarity in a shared multimodal embedding space. A high $C_f$ identifies a "Phantom Affordance", i.e. a strong architectural signal that creates a spatial promise the physical environment fails to keep.

Mapping these divergence points generates \textbf{Cognitive Friction Heatmaps} that diagnose semiotic ambiguity. This treats hallucinations not as failures, but as \textbf{Latent Design Requirements}. 
Consequently, the design strategy shifts from correcting the agent’s perception to aligning the environment’s cues (e.g. lighting, texture, and transparency) with the agent's expectation.

\begin{figure}[h!]
    \centering
    \includegraphics[width=1\linewidth]{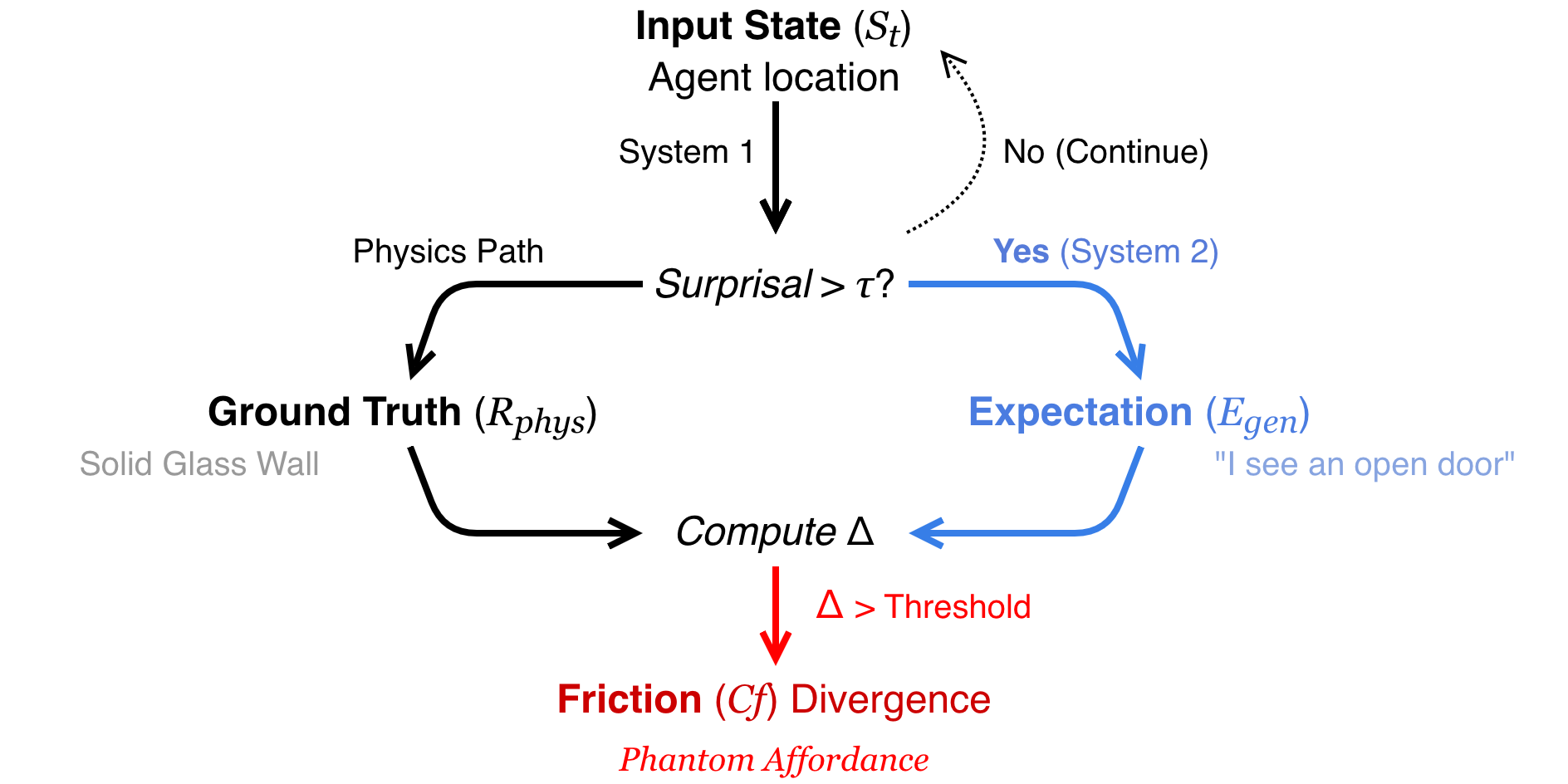}
    \caption{\textbf{The Cognitive Friction Pipeline.} The simulation operates on a physics autopilot (System 1) until prediction error (surprisal) exceeds a threshold $\tau$ \citep{Zacks2007EventSegmentation}. This triggers the VLM (System 2) to generate an expectation ($E_{gen}$). The semantic divergence ($\Delta = 1 - sim$) from ground truth ($R_{phys}$) quantifies the Cognitive Friction ($C_f$); high friction reveals a "Phantom Affordance".}
    \label{fig:pipeline}
\end{figure}

\section{Ethical Implications: From Optimization to Orchestration}


Although the technical capacity for affect-aware simulation exists to some extent, the theoretical framework for interpreting these "hallucinated" user experiences remains underdeveloped. 
We argue that generative AI predictions of affect and cognitive load can threaten cognitive integrity by enabling manipulative, data-driven "nudging" in environments that know users too well
~\cite{Helbing2019}.

Building on Floridi's ethics of artificial intelligence \citep{Floridi2023TheOpportunities} and opposing to technocratic visions, we propose that these simulations must not only optimize for efficiency 
but for cognitive clarity, i.e. ensuring that the environment's adaptive behavior  is transparent to the individual. To this end, we propose a new HCI paradigm, that shifts from "Environment Optimization" to "Cognitive Orchestration". 
This framework ensures that AI-driven spatial adaptations preserve human autonomy rather than reducing occupants to predictable data points. 

\textbf{Design Principles for Cognitive Orchestration:}

\begin{enumerate}
    \item \textbf{Interpretability:} $C_f$ heatmaps must be legible to non-experts, enabling occupants to understand why spaces feel disorienting or stressful.
    \item \textbf{Opt-Out Mechanisms:} Adaptive environments should provide manual overrides, preserving individual agency over spatial experience.
    \item \textbf{Demographic Equity:} Simulation training data must represent diverse populations to prevent optimizing environments for dominant demographic groups.
    \item \textbf{Audit Trails:} All simulation parameters, training data sources, and design decisions informed by $C_f$ metrics should be documented and contestable.
\end{enumerate}

\section{Discussion: Validation and Limitations}

\textbf{Contributions.} This paper presents three contributions:
(1) a conceptual shift from chronological simulation to episodic spatial reasoning aligned with human memory structures;
(2) a formalization of generative hallucinations as a cognitive friction metric for diagnosing semiotic ambiguity in built environments;
and (3) a human-centered framework of cognitive orchestration that reframes AI-driven environmental simulation as an ethical design practice rather than an optimization problem.


\textbf{Generalization Concerns.} Current implementations assume Western spatial semiotics. Cultural variation in architectural affordances (e.g., threshold meanings, proxemic norms) may limit cross-cultural applicability. Future work must diversify training data across cultural contexts and validate $C_f$ metrics in non-Western environments. Moreover, computational costs may constrain real-time applications, and the framework's dependence on specific VLM architectures raises concerns about metric stability as models evolve.

\textbf{Empirical Validation.} This work formalizes cognitive friction metrics at a computational level. Empirical validation with human behavioral data remains future work, with planned pilot deployments in hospital and transit environments, among others, to test relations between predicted friction hotspots and user experience.

\textbf{Ethical Guardrails.} While $C_f$ metrics identify problematic friction, their application requires a governance framework preventing misuse for behavioral manipulation. We advocate for participatory design processes where affected communities audit simulation parameters and contest findings. The distinction between revealing cognitive hazards (ethical) and engineering compliance (unethical) must remain central.

\textbf{From Diagnosis to Generation.} While current work focuses on diagnostic applications of $C_f$ metrics, an inverse approach could leverage hallucinations generatively: starting with narrative prompts (e.g., "a high-trust negotiation space") rather than fixed geometries, aggregating agent hallucinations as latent design requirements, and iteratively generating spatial configurations that fulfill these emergent expectations.
This means treating geometry as a "self-fulfilling prophecy" aligned with cognitive intent rather than a predetermined constraint.

\section{Conclusion}

Agentic environmental simulations represent a paradigm shift in modeling human-environment interactions. By treating occupants as cognitive agents rather than physical particles, employing episodic spatial reasoning aligned with human memory structures, and formalizing generative errors as diagnostic tools through the cognitive friction metric, we enable prediction of experiential qualities previously accessible only through a post-occupancy evaluation. 

The $C_f$ metric and the concept of Phantom Affordances provide actionable frameworks for identifying where built environments fail to support cognitive intent. However, this technical capability demands ethical frameworks prioritizing cognitive integrity over behavioral optimization. Our principle of cognitive orchestration ensures AI-driven spatial design enhances rather than undermines human autonomy, affective clarity, and cognitive sovereignty.

As these technologies mature, the architectural discipline faces a choice: use predictive cognitive modeling to manipulate behavior or to orchestrate environments that respect the full complexity of human spatial experience. We advocate for the latter, grounded in transparency, equity, and respect for human autonomy.


\bibliographystyle{ACM-Reference-Format-lim}
\bibliography{sample-base}

\appendix

\end{document}